\documentclass[11pt,twoside]{article}
\usepackage{latexsym} 
\usepackage{amssymb,amsbsy,amsmath,amsfonts,amssymb,amscd}
\usepackage{graphicx}
\usepackage{hyperref}  
\usepackage{xcolor} 
\newcommand {\wt}[1] {{\widetilde #1}}
\newcommand{\commentout}[1]{}
\newcommand{\R}{\mathbb{R}}

\newcommand {\al} {\alpha}
\newcommand {\e}  {\varepsilon}

\newcommand {\sg} {\sigma} 
\newcommand {\vp} {\varphi} 

\newcommand {\Chi} {{\bf \raise 2pt \hbox{$\chi$}} }
\newcommand {\f}   {\frac}
\newcommand {\p}   {\partial}
\newcommand{\dis}{\displaystyle}
\newcommand {\proof} {\noindent {\bf Proof}. }
\newcommand{\beq}{\begin{equation}}
\newcommand{\eeq}{\end{equation}}
\newcommand{\bea} {\begin{array}{rl}}
\newcommand{\eea} {\end{array}}
\newcommand{\bepa}{\left\{ \begin{array}{ll}}
\newcommand{\eepa} {\end{array}\right.}
\newtheorem{theorem}{Theorem}[section]
\newtheorem{lemma}[theorem]{Lemma}

\newtheorem{proposition}[theorem]{Proposition}
 
\newcommand{\qed}{{ \hfill
                       {\unskip\kern 6pt\penalty 500 \raise -2pt\hbox{\vrule\vbox to 6pt{\hrule width 6pt
                       \vfill\hrule}\vrule} \par}   }}
\title{Distributed synaptic weights in a LIF neural network \\ and learning rules}

\author{
Beno\^ \i t Perthame\thanks{Sorbonne Universit\'es, UPMC Univ Paris 06, CNRS UMR 7598, Laboratoire Jacques-Louis Lions, Inria \'Equipe MAMBA, 4, place Jussieu 75005, Paris, France, Email:~benoit.perthame@upmc.fr} 
\and Delphine Salort\thanks{Sorbonne Universit\'es, UPMC Univ Paris 06, CNRS UMR 7238 , Laboratoire de Biologie Computationnelle et quantitative,  4, place Jussieu 75005, Paris, France, Email:~delphine.salort@upmc.fr} 
\and Gilles Wainrib\thanks{Ecole Normale Superieure France, D\'epartement d'Informatique, \'equipe DATA, Paris, France, Email:~gilles.wainrib@ens.fr} 
}

\date{\today}

\begin{document} 
\maketitle
\pagestyle{plain}
\pagenumbering{arabic} 


\begin{abstract} Leaky integrate-and-fire (LIF) models are mean-field limits, with a large number of neurons, used to describe neural networks. We consider inhomogeneous networks structured by a connectivity parameter (strengths of the synaptic weights) with the effect of processing the input current with different intensities. 
\\

We first study the properties of the network activity depending on the distribution of synaptic weights and in particular its discrimination capacity. Then, we consider simple learning rules and determine the synaptic weight distribution it generates. 
We outline the role of noise as a selection principle and the capacity to memorized a learned signal.
\end{abstract}

\bigskip


%
\section{Introduction}
\label{sec:intro}

Learning and memory are essential cognitive functions which are supported by the subtle mechanisms of synaptic plasticity \cite{martin2000synaptic}. Since the seminal work of Hebb \cite{hebb1949organization}, one of the key challenges in theoretical neuroscience and artificial intelligence is to understand the consequences of various learning rules on the organization of neural networks and on the way they process and memorize information. 

Despite several theoretical works on this topic \cite{gerstner2002spiking,dayan2001theoretical,gerstner2002mathematical,galtier2013biological,galtier2012multiscale} it remains difficult to investigate learning processes using macroscopic models with infinite number of neurons because these usually assume a form of homogeneity, either under uniform synaptic weights assumptions leading to McKean-Vlasov limits \cite{dai1996mckean,bertini2010dynamical,Demasietal,delarue2015global,BossyFaugerasTalay} or under random connectivity models leading to dynamical spin-glass limits \cite{sompolinsky1988chaos,arous1995large}. In contrast, when studying synaptic plasticity and learning, one needs to describe all the connections between each pair of neurons, hence breaking the homogeneity usually necessary to derive such macroscopic limits. 

In this article, we propose a way to circumvent this difficulty by introducing a mathematical model describing a macroscopic population of leaky integrate-and-fire neurons (LIF in short), which are interacting through a mean-field variable 
 and where learning rules are governed by the mean activity of the network. More precisely, 
in contrast with previous works on such 
questions, we consider a model where neuronal subpopulations interact with the mean-field through a heterogeneous distribution of synaptic weights values: in other words, each subpopulation sees the mean-field through a different lens. Instead of considering activity-dependent changes in the pairwise synaptic weights between neurons, we consider that each subpopulation receives a weighed version of the overall network activity, and that the associated weights can be dynamically modified according to a specific rule. Based on this heterogeneous model, we are therefore able to integrate a learning rule term in the equation, for the first time in the class of macroscopic LIF equations. One particular instance of such learning rule corresponds to the idea of Hebbian learning, in the sense that the connection between a given subpopulation and the mean-field is strengthen if both have a correlated activity and is weakened otherwise.

The introduction of this mathematical framework supports the investigation of several questions inspired by the seminal work of Hopfield \cite{hopfield1982neural}, which are answered, at least partially, in this article:
\begin{enumerate}
\item For a given pattern of steady-state neural activity, can we always find a heterogeneous synaptic weight distribution that generates such activity?
\item What is the equilibrium synaptic weight distribution according to the learning rule and to the external signal?
\item Is the system able to remember which external signal was presented during the learning phase?
\end{enumerate}

We  present in  Section~\ref{sec:mfe} the different mathematical models that we consider in order to study the effect of a mean field learning rule on coupled neural networks governed by Noisy LIF models structured by a connectivity parameter. In  Section~\ref{sec:nonlinear}, we introduce some material about the possible stationary solutions of these models without learning rule, in particular we study existence and uniqueness. This material will be use throughout the paper. 
Next, we focus our study on qualitative properties on the input-output map and on the learning rules. 
In Section~\ref{sec:given}, we prove that our model can produce a large class of output signals by an appropriate choice of  the distribution of connectivity. In Section~\ref{sec:discrimination},  we show the ability of our model to differentiate different inputs via a discrimination property: we prove that we cannot obtain the same  output signal considering two different  input signals and, given two different inputs,  we give an estimate of the difference between the two possible output signals  associated to two different inputs.   The last sections are devoted to the study with the learning rule. In Section~\ref{sec:connectivity}, we address the full system with learning and we describe  possible equilibrium connectivity distributions and prove non-uniqueness. We  propose in Section~\ref{sec:noise} a selection principle by adding noise in the learning stage. In Section~\ref{sec:LTPR}, we describe the ability of such system to memorize learned signals by discriminating easily new incoming inputs after learning.

\section{Mathematical models of a mean field learning rule}
\label{sec:mfe}

\textcolor{black}{It is standard to describe homogeneous neural networks by LIF models. In order  to  study the impact of heterogeneity and of mean field learning rules, we introduce a mathematical model  for coupled neural networks. To this end, we firstly explain the equations which describe the activity of a macroscopic population of LIF neural networks when they interact through they mean  activity. Secondly, we present how mean field learning rules  may be included in these equations. } 

\subsection{ \textcolor{black}{Structured Noisy LIF model}}

We consider a heterogeneous population of  homogeneous neural networks structured by their  synaptic weights $w \in (-\infty,+\infty)$, negative sign stands for inhibitory neurons and positive sign for excitatory neurons.  
\textcolor{black}{We have  chosen to use a signed parameter $\sg$, instead of a system of two equations for excitatory and inhibitory neurones,  because this leads to a simpler formalism and avoids boundary conditions in $w=0$.
}
We assume that each homogeneous subpopulation, with synaptic weight $w$, is governed by the classical mean field noisy integrate and fire equation, widely used for large neural networks  \cite{BrHa, brunel,BrGe}.  Moreover, we assume that the subpopulations interact via the total firing rate $\bar N(t)$ defined as  the mean activity for all the subpopulations. Setting $v$ the value of the action potential, $V_F >V_R$ the values of the firing and reset potentials, we consider the classical equation
\begin{equation}\label{eqsc:1}
\frac{\partial p}{\partial t} + \frac{\partial}{\partial v} \left[\big(- v + I(t,w) + w \sigma(\bar N(t)) \big) p \right]
- a \frac{\p^2 p}{\p  v^2} = N(w,t)  \delta(v-V_R), 
 \end{equation}
with the boundary and initial conditions  
\begin{equation} \label{eq:2}
p(V_F,w,t)=0, \qquad p (- \infty,w,t)=0,  \qquad p(v,w,0)=p^0(v,w).
\end{equation}
The solution  $p(v,w, t)$ defines the probability to find a neuron at potential $v$ with a synaptic weight $w$, the coefficient $a$ represents the synaptic noise which we assume to be constant 
and $I(t, w)$ is the input signal which strength is possibly modulated by the synaptic weights. 
The subnetwork activity $N(w,t)$ and  total activity $\bar N(t)$ are defined as 
\begin{equation} \label{eq:3}
N(w,t):= -a \frac{\p p}{\p v} (V_F,w,t) \geq 0, \qquad \bar N(t)= \int_{-\infty}^\infty  N(w,t) dw .
\end{equation}
The function $\sg( \cdot)$ represents the response \textcolor{black}{ of the network to the total activity. We  use either $\sigma(N)=N$, or  the following class with saturation} 
\begin{equation}
\sigma \in \mathcal{C}^{2}(\R^+; \R^+), \qquad \sg_M=  \max \sg(\cdot) < \infty, \qquad \sigma' \geq 0.
\label{as:sigma}
\end{equation}
 \textcolor{black}{We recall some mathematical properties of distributional solutions of Equation~\eqref{eqsc:1}--\eqref{eq:3} which were studied in \cite{CCP, CGGS, CPSS}. There, some existence, uniqueness and long time behaviour results are established for distributional solutions.
 }%
\textcolor{black}{ For excitatory networks, solutions can blow-up in finite time, as discovered in~\cite{CCP}, when $\sg(N) =N$.
The saturation assumption~\eqref{as:sigma}  prevents blow-up}, a phenomena which also appears if one uses the activity dependent noise $a:= a_0+a_1 N, (see $~\cite{CPSS}), or if a refractory state is included~\cite{CaPe}. 
 \textcolor{black}{ In the inhibitory case, blow-up  never occurs when noise is independent of the activity~\cite{CGGS,CPSS} and solutions are globally bounded. This holds even for $\sg(N) =N$ and  assumption \eqref{as:sigma} is not fundamental in the inhibitory case.  Then, the main open problem which remains is to prove the long time convergence; only small perturbations of the linear case  are treated so far.}

The initial data  $p^0(v,w)\geq 0$ is a probability density and a basic property of the above LIF model is that
\begin{equation} \label{eq:proba}
\int_{-\infty}^\infty \int_{-\infty}^{V_F} p(v,w,t)\,dv dw = \int_{-\infty}^\infty  \int_{-\infty}^{V_F} p^0(v,w)\,dv dw=1. 
\end{equation}
 We finally define the probability density of neural subnetworks with synaptic weight $w$ by 
\begin{equation} \label{eq:connectivity_distr}
H(w,t) = \int_{-\infty}^{V_F} p(v,w,t)\,dv, \qquad \int_{-\infty}^\infty  H(w,t) dw =1.
\end{equation}
 Let us mention that, so far, the function $H$ is independent of time because in Equation~\eqref{eqsc:1}, the distribution of synaptic weights is fixed.  
 Moreover, with this distribution $H$, an input signal $I(w)$ is stored as a {\em normalized output signal} which we define thanks to  the network activity as 
\beq
S(w)= \f{N(w) }{\bar N} \geq 0, \qquad \int_{-\infty}^{+\infty}  S(w) dw=1.
\label{def:output}
\eeq
The normalization is due to the size of the network normalized by $\int H(w)dw =1$ which induces a limitation of possible outputs.
\newline
 Let us now include some learning rules that may modify this distribution.
 
\subsection{\textcolor{black}{Models with mean field learning rules.}}

\noindent  \textcolor{black}{ Next, we introduce some learning rules in order to  modulate} the distribution of synaptic weights $H$ and allow the network to recognize some given input signals~$I$ by choosing an appropriate heterogeneous synaptic weight distribution $H$ adapted to the signal $I$.  To this end, we have chosen learning rules inspired from the seminal Hebbian rule  which essentially consists in assuming that the strength of weights $w_{ij}$ between two neurons $i$ and $j$ increases when the two neurons have high activity simultaneously. For $M$ neurons in interactions, the classical Hebbian rule relates the weights to the activity $N_i$ of the neuron $i$
$$
\frac{d}{dt}w_{ij}=  k_{ij} N_iN_j -w_{ij} , \quad  1\leq i , \; j \leq M. 
$$

 In our context, we assume that the subnetworks interact only via the total firing rate $\bar N$,  with synaptic weights described with a single parameter $w$, not a matrix. Hence,  we cannot generalize directly the Hebbian learning rule and we give the following interpretation. All the subnetworks parametrized by $w$ may modulate their intrinsic  synaptic weight  $w$ with respect to a function  $\Phi$ which depends on the intrinsic activity $N(w)$ of the network parametrized by $w$ and of the total activity of the network $\bar{N}$.   Then, the proposed generalization of the Hebbian rule consists in choosing 
 \beq
 \Phi(N(w),\bar{N})=\bar{N} N(w) K(w),
 \label{hebbian}
 \eeq 
 where $K(\cdot)$ represents the  learning  strength of the subnetwork with synaptic weight $w$.

 Adding the above choice of  learning rule, we obtain the following equation
\begin{equation}\label{eq:1} \left\{\begin{array}{l}
\frac{\partial p}{\partial t} + \frac{\partial}{\partial v} \left[\big(- v + I(w) +w \sigma(\bar N(t)) \big) p \right]
+ \varepsilon  \frac{\partial}{\partial w} \left[\big( \Phi  - w \big) p \right]
- a \frac{\p^2 p}{\p  v^2} = N(w,t)  \delta(v-V_R), 
\\[5pt]
N(w,t) : =  -a \frac{\p p}{\p v} (V_F,w,t) \geq 0, \qquad \bar N(t)= \dis \int_{-\infty}^\infty  N(w,t) dw, 
\end{array} \right.
 \end{equation}
with the boundary and initial  conditions 
$$
p(V_F,w,t)=0, \qquad p (-\infty,w,t)=0, \qquad p (v, \pm \infty,t)=0, \qquad p (v,w,0)=p^0(v,w).
$$
Here, $\e$ stands for a time scale which takes into account that learning is slower than the normal activity of the network, and $ \Phi= \Phi \big(N(w,t), \bar N(t) \big)$ represents  the learning rule. Notice that a desirable property is that the flus $\Phi-w$ is inward which occurs for instance when $\Phi$ is bounded or sub-linear at infinity. 
Several  direct extensions  of the Hebbian rule are possible for example
$$
\Phi = N(w,t) \int K(w,w') N(w',t) dw',
$$
or inspired from STDP rule \textcolor{black}{(spike timing dependent plasticity, see \cite{gerstner2002spiking} for instance),}   where post- and pre-synaptic spike times are compared, we may  choose
$$
\Phi \big(N(w,t), \bar N(t) \big)  =  \Phi\big(\big(N(w,t)*_t g \bar N(t) - N(w,t) \bar N(t)*_t g \big) .
$$
Here $g(t) =0$ for $t<0$ and $g(t)=e^{-t/\tau}$ for $t>0$.



\section{ Stationary solution of the nonlinear problem  without learning rule} 
\label{sec:nonlinear}

\textcolor{black}{Throughout this paper we use some material about the possible stationary solutions of Equations~\eqref{eqsc:1}--\eqref{eq:3}  submitted to a given  input  signal $I(w)$. These stationary states are defined through the equation} 
\begin{equation}\label{eq:nifstst1}
 \frac{\partial}{\partial v} \left[\big(- v + I(w)+ w \sg(\bar N) \big) P(v,w) \right]
- a \frac{\p^2 P(v,w)}{\p  v^2} = N(w)  \delta(v-V_R), 
 \end{equation}
with the boundary conditions 
\begin{equation} \label{eq:nifstst2}
P(V_F,w)=0, \quad P (-\infty,w)=0,   \quad \text{and } \quad N(w)= -a \frac{\p P}{\p v} (V_F,w) \geq 0. 
\end{equation}
We recall that the nonlinearity is driven by the network total activity  defined as
\begin{equation} \label{eq:nifstst3}
\bar N= \int N(w) dw ,
\end{equation}
and that we assume a normalization~\eqref{eq:proba} which is written 
\begin{equation} \label{eq:nifstst4}
\int_{-\infty}^{V_F} P(v,w) dv  = H(w), \qquad \int H(w) dw =1.
\end{equation}

Our  first result is the

\begin{theorem} [Existence of stationary states]
We assume \eqref{as:sigma}, give the input signal $I\in  L^\infty (\R)$ and the synaptic weight distribution $H(w)$ normalized to $1$ such that there exists $\varepsilon >0$ with 
\beq
\int_0^\infty  w^2 H(w)  dw < \infty.
\label{as:H}
\eeq 
Then, there is at least one solution of \eqref{eq:nifstst1}--\eqref{eq:nifstst4}.
\\
In the case of inhibitory network, that is $supp  (H )\subset (-\infty, 0]$, for all $\sigma\in  \mathcal{C}^2$, the solution is unique. 
\label{thm:existence}
\end{theorem}

 Let us mention that for a single $w$, semi-explicit formula are available, see \cite{CCP} for instance, and  the stationary states are not necessarily unique in the excitatory case. 
\\

\proof Our approach is  to solve the nonlinear problem using a fixed point argument on the  value $\bar N$. Being given $\bar N$, $I(w)$ and $H(w)$, we consider the linear problem where $w$ is a parameter (we do not repeat the boundary conditions)
\begin{equation}\label{eq:linear} 
\bepa
 \frac{\partial}{\partial v} \left[\big(- v + I(w) + w \sg\big(\bar N\big) \big) Q_{\bar N,I} \right]
- a \frac{\p^2 Q_{\bar N,I}}{\p  v^2} = N_{\bar{N}, I}(w)  \delta(v-V_R), 
\\[15pt]
Q_{\bar N,I}(V_F,w )= 0, \quad N_{\bar{N}, I}(w) = - a \frac{\p Q_{\bar N,I}(V_F,w)}{\p  v},  \quad \int_{-\infty}^{V_F} Q_{\bar N,I}(v,w)dv=1.
\eepa
\end{equation}
\textcolor{black}{ Let $\psi : \R^+ \to \R^+$ defined by 
$$\psi(\bar{N})= \int_{-\infty}^{+\infty} N_{\bar{N}, I}(w)H(w) dw.$$
Then, we obtain a solution of \eqref{eq:nifstst1}--\eqref{eq:nifstst3} if and only if $\bar{N}$ is a fixed point of the application $\Psi$, that is  }
\begin{equation}\label{eq:FPC}
\int_{-\infty}^{+\infty} N_{\bar{N}, I}(w)H(w) dw= \bar N, \qquad \hbox{and then} \; P(v, w) =H(w)Q_{\bar N,I}(v,w) .
\end{equation}
To prove the existence of such a fixed point, we need a careful analysis of the mapping $\bar N \mapsto N_{\bar{N}, I}(w)$, which we perform in the next subsection, adding many properties that will be used later on. The conclusion of the proof is given afterwards.

\subsection{Main properties of  $N_{\bar{N}, I}(w)$ in \eqref{eq:linear}}
\label{sec:propN}

Being given $\bar N$, the  linear stationary state Equation~\eqref{eq:linear}, because it is solved $w$ by $w$,  is a standard equation and solutions form a one dimensional vector space (the eigenspace for the eigenvalue $0$) according to the Krein-Rutman theorem \cite{DautrayLions6}. Uniqueness is enforced thanks to the normalization as a probability.
\\

 Integrating Equation~\eqref{eq:linear}, we obtain that a solution satisfies 
\beq
\big(- v +I (w)+ w \sg(\bar N) \big) Q_{\bar N,I}(v,w) - a \frac{\p Q_{\bar N,I}(v,w)}{\p  v} = 
\begin{cases}
\; 0 \qquad \qquad  &\text{for }\; v<V_R,
\\[5pt]
\; N_{\bar{N}, I}(w)  \qquad  &\text{for }\; v> V_R,
\end{cases}
\label{equationN}
\eeq 
with $N_{\bar{N}, I}(w)$ to be found such that  $ \int_{-\infty}^{V_F}Q_{\bar N,I}(v,w) dv=1$. Hence, the solution is explicitly given by 
\beq
Q_{\bar N,I}(v,w)  = 
\begin{cases}
\f{1}{Z_{\bar N,I}(w)} e^{-\f{(v- I(w)  - w \sg(\bar N))^2}{2a}} \dis \int_{V_R}^{V_F}  e^{\f{(v' -I(w) - w \sg(\bar N))^2}{2a}} \,dv'  \qquad &\text{for }\; v<V_R,
\\[15pt]
\f{1}{Z_{\bar N,I}(w)} e^{-\f{(v -I(w) - w \sg(\bar N))^2}{2a}} \dis \int_{v}^{V_F}  e^{\f{(v' -I(w)- w \sg(\bar N))^2}{2a}} \,dv'   \qquad  &\text{for }\; v> V_R,
\end{cases}
\label{solutionQ}
\eeq
with
$$
 Z_{\bar N,I}(w) = \dis \int_{-\infty}^{V_F}  \dis \int_{v'= \max(V_R,v) }^{V_F} e^{\f{(v'  -I(w)- w \sg(\bar N))^2- (v -I(w) - w \sg(\bar N))^2}{2a}} \,dv'  dv
$$
which is also written under the more convenient form
\beq
 Z_{\bar N,I}(w) = \dis \int_{-\infty}^{V_F}  \dis \int_{v'= \max(V_R,v) }^{V_F} e^{\f{(v' - v). [v'+v -2I(w)-2 w \sg(\bar N)]}{2a}} \,dv'  dv .
\label{eq:Z}
\eeq
Because of the boundary condition $Q_{\bar N,I}(V_F,w) =0$, an immediate consequence of \eqref{equationN} is the relation 
\beq
N_{\bar{N}, I}(w) = \f{a}{Z_{\bar N,I}(w)}= - a \frac{\p Q_{\bar N,I}(V_F,w)}{\p  v}  . 
\label{ifststs:Zdef}
\eeq
 
Throughout the paper, we use properties of the function $Z_{\bar N,I}(w)$ which we state in the following lemma.
\begin{lemma}\label{Linearproblem} Given $I\in L^\infty(\R)$ and $\bar N>0$, the unique solution $Q_{\bar N,I}(v,w) >0$ of  Equation~\eqref{eq:linear}, defined by \eqref{solutionQ}-\eqref{eq:Z}, satisfies the following estimates. There is a constant $C(a, V_R, V_F)$  such that
\textcolor{black}{
\beq 
C \min\left( \f {1}{\| I \|_{L^\infty} +|w|_+ \sigma(\bar N) }, (V_F-V_R) \right)^2
  \leq Z_{\bar N,I}(w) \leq C e^{\f{\big(\| I \|_{L^\infty} +| w|_- \sg(\bar N) \big)^2}{a}},
\label{ifststs:Zupper}
\eeq}
\beq 
  \lim_{w \to +\infty}Z_{\bar N,I}(w) =0, \quad   \lim_{w \to -\infty}Z_{\bar N,I}(w) =+\infty,  \quad  \inf_{\bar N} Z_{\bar N,I}(w) >0, \quad  \forall w \in \R,
\label{ifststs:Zmin}
\eeq
\beq
\forall w \leq 0 ,   \quad \p_{\bar N} Z_{\bar N,I}(w) \geq 0 \quad  \hbox{ and } \quad  \forall w \geq 0, \quad \p_{\bar N} Z_{\bar N,I}(w) \leq 0  ,
\label{ifststs:Zprime}\eeq
\textcolor{black}{ 
\beq
\p_{w} Z_{\bar N,I}(w) \leq 0, \quad  \hbox{ if  $I'(\cdot) \geq 0$.}
\label{ifststs:Zprimew} 
\eeq
Moreover, when  $\sigma(N)= N$, the following estimates holds}
 \begin{equation}\label{eqinfty}
\lim_{-w\bar N  \to +\infty}Z_{\bar N,I}(w) =+\infty ,
\end{equation}
\beq
 \partial_{\bar N \bar N}^{2} Z_{\bar N,I} (w)>0, \quad \forall w\leq 0. 
 \label{Zsecond} \eeq
\end{lemma}

\noindent {\bf Proof of Lemma \ref{Linearproblem}.}
\textcolor{black}{ We first prove inequality \eqref{ifststs:Zupper}. We set $A= \| I \|_{L^{\infty}} +| w|_- \sg(\bar N) $ where $|w|_-= -\min(0,w)$.}  As 
$$
 e^{ \frac{2(v'-v)(-I(w)- w  \sigma (\bar{N}))}{2a} }  \leq  e^{ \frac{ 2(V_F-v) A}{2a} }, \quad  \hbox{ for }   V_F \geq v' \geq v,
$$
we obtain,  with formula   \eqref{eq:Z}, that there exists a constant $C$ such that
$$
 Z_{\bar N,I}(w) \leq  \dis \int_{-\infty}^{V_F}  \dis \int_{v'= \max(V_R,v) }^{V_F} e^{\f{|v'|^2 -|v |^2 + 2(V_F-v) A}{2a}} \,dv'  dv
 \leq C   \int_{-\infty}^{V_F}  e^{\f{-|v|^2+ 2(V_F-v) A}{2a}} \ dv.
 $$
Therefore, we conclude the upper bound with
 $$  
Z_{\bar N,I}(w) \leq C  e^{ \frac{2V_F A}{2a}} \int_{-\infty}^{V_F}  e^{\f{-|v|^2-2vA}{2a}} \ dv
=  C  e^{ \frac{V_F^2+ A^2+A^2}{2a}} \int_{-\infty}^{V_F}  e^{\f{-|v|^2-2vA-A^2}{2a}} \ dv.  
$$
 For the lower bound, we set
$$
|w|_+=\max(0,w) \hbox{  and  } \widetilde{A}=  \| I \|_{L^\infty} +|w|_+ \sigma(\bar N) ,
$$
and conclude that
$$  
e^{ -  \frac{(V_F-v) \widetilde{A}}{a}  }  \leq e^{ \frac{2(v'-v)(-I(w)- w  \sigma (\bar{N}))}{2a} } \quad  \hbox{ for }  \; V_F \geq v' \geq v .
$$
Inserting this lower bound in formula \eqref{eq:Z}, we deduce that there exists a constant $C(a,V_R,V_F)$ such that 
$$
 Z_{\bar N,I}(w) \geq  \int_{-\infty}^{V_F} (V_F- \max(V_R,v)) e^{\f{- |v|^2  -2 (V_F-v) \widetilde{A}}{2a}}dv
 \geq C \dis \int_{V_R}^{V_F} (V_F-v) e^{\f{  - (V_F-v) \widetilde{A}}{a}}   dv = C  \dis \int_{0}^{V_F-V_R} z e^{\f{  - z \widetilde{A}}{a}} dz .
$$
We deduce that there exists a constant $C(a,V_R,V_F)$ such that 
$$ 
Z_{\bar N,I}(w) \geq  \dis C \min(\f {1}{\widetilde{A}^2}, (V_F-V_R)^2 ),   
$$
which ends the proof of estimate \eqref{ifststs:Zupper}. 
\\
Next, we prove the inequality \eqref{ifststs:Zmin}.  We have, for all $w \in \R$, 
\beq \bea
 \dis  \int_{-\infty}^{V_F}  \int_{v'= \max(V_R,v) }^{V_F} &e^{\f{(v' - v). (v'+v -2 \|I\|_{L^{\infty}}-2 w \sg(\bar N))}{2a}} dv'  dv  \leq  Z_{\bar N,I}(w)   
  \\[15pt]
&\leq   \dis\int_{-\infty}^{V_F}  \dis \int_{v'= \max(V_R,v) }^{V_F} e^{\f{(v' - v). (v'+v +2 \|I\|_{L^{\infty}}-2 w \sg(\bar N))}{2a}}dv'  dv .
   \eea
\label{ifststs:Z0}
\eeq
Because  $\sigma >  0$ for $\bar{N} \neq 0$, we have almost everywhere in $v$ and $v'$
$$ 
\lim_{w \to -\infty} e^{-2w  \sg(\bar N)(v'-v)}= +\infty \quad \hbox{ and }  \quad  \lim_{w \to +\infty}e^{-2w  \sg(\bar N)(v'-v)}   = 0.
$$  

To prove the estimate \eqref{ifststs:Zprime}, we differentiating  the explicit formula of  $ Z_{\bar N,I}(w)$ with respect to $\bar N$. We  obtain that
$$
\p_{\bar N} Z_{\bar N,I}(w)  = - w \sg'(\bar N)  \dis \int_{v= -\infty}^{V_F}\int_{v'= \max(V_R,v) }^{V_F} \f{v'-v}{a} e^{\f{(v' -I(w)- w \sg(\bar N))^2}{2a}}  e^{-\f{(v - I(w)-w\sg(\bar N))^2}{2a}} dv \, dv' ,  
$$
which directly gives \eqref{ifststs:Zprime}.
\\

\textcolor{black}{To prove \eqref{ifststs:Zprimew},  we observe that, when $I'(w) \geq 0$,
$$
\partial_w Z_{\bar N,I}(w) = -(I'(w)+ \sigma (\bar{N}))   \dis \int_{v= -\infty}^{V_F}\int_{v'= \max(V_R,v) }^{V_F} \f{v'-v}{a} e^{\f{(v'-v)(v'+v  - 2 I(w)- 2w \sg(\bar N))}{2a}}  dv \, dv'  \leq 0.
$$}
Next, when  $\sigma(N)=N$, we have 
$$ 
Z_{\bar N,I}(w)  \geq  \dis \int_{v=w \bar N}^{w \bar N +1} e^{-\f{(v -I(w) - w \bar N)^2}{2a}} \dis \left( \dis \int_{v'=\max(v,V_R)}^{V_F}  e^{\f{(v' -I(w) - w \bar N)^2}{2a}} \,dv' \right) dv .
$$
Therefore, we obtain estimate \eqref{eqinfty} because for $v  \in (w \bar N,w \bar N+1)$,  we have
$$
e^{-\f{(v -I(w) - w \bar N)^2}{2a}} \geq e^{-  \frac{(\|I\|_{L^\infty}+2)^2}{2a}},\quad  \hbox{so that} \quad 
Z_{\bar N,I}(w)  \geq e^{-  \frac{(\|I\|_{L^\infty}+2)^2}{2a}} \int_{v'=V_R}^{V_F}  e^{\f{(v' -I(w) - w \bar N)^2}{2a}} dv' .
$$
Finally, the inequality \eqref{Zsecond}  follows from 
$$
\partial_{\bar N \bar N}^2 Z_{\bar N,I}= \frac{w^2}{a^2}   \dis \int_{v= -\infty}^{V_F}   \int_{v'=\max(v,V_R)}^{V_F} (v'-v)^2  e^{-\f{(v -I(w) -w  \sg(\bar N))^2}{2a}} e^{\f{(v' -I(w) - w  \sg(\bar N))^2}{2a}}dv'dv >0.
$$
The proof of  Lemma  \ref{Linearproblem} is complete. 

\hfill $\square$

\subsection{Conclusion of the proof of Theorem \ref{thm:existence}}
\label{sec:concl_exist}

We come back to the fixed point equation \eqref{eq:FPC}. With the above notations, it is restated, using the quantity ${Z}_{\bar N, I}(w)$ defined by~\eqref{eq:Z}, as a fixed point of the function $\psi: \R^+ \to \R^+$
\beq
\psi(\bar N):= a \int_{-\infty}^{+\infty} \f{H(w)}{{Z}_{\bar N, I}(w)} dw= \bar N . 
\label{eq:FPCbis}
\eeq
We have $\psi(0)>0$.
\newline
In the inhibitory case, when supp$(H)\subset \R^-$, we have $\psi'(\cdot) <0$ thanks to~\eqref{ifststs:Zprime}, and thus there is a unique fixed point. 
\newline
When excitatory weights are considered, as 
$$
  \lim_{w \to +\infty}  \frac{1}{\bar{Z}_{\bar N, I}(w)} =0,
$$
we have to impose an additional assumption on $H$ to control $\psi( \cdotp)$. For this purpose, using estimate \eqref{ifststs:Zupper}, thanks to the assumption~\eqref{as:H}, we know that $\psi(\cdot)$ remains bounded and  hence, there is at least one fixed point by continuity.
\hfill $\square$

\section{Output signals  induced from  a synaptic weight distribution} 
\label{sec:given}

As a first property of the network properties, we aim at identifying which possible steady states output activities $N(w)$ or signal $S(w)$ can be  generated by the network by varying the synaptic weight distribution~$H$. 

We prove that any  nonnegative normalized output signal  $S \in L^1(\R)$, with fast decay at $-\infty$, can be, up to  a multiplicative constant,  reproduced by a stationary state of Equation~\eqref{eq:nifstst1} for a  well chosen synaptic weight distribution $H$.

%
\begin{theorem} [Relation output signal to synaptic weights]
We assume \eqref{as:sigma} and give the input signal $I\in  \mathcal{C}^1_{\rm b} (\R)$ and the output signal $ S \geq 0$ normalized with $\int_{-\infty}^{+\infty} S(w) =1$ and satisfyning $\int_{-\infty}^{0} e^{- \gamma w} S(w) < \infty$ with $\gamma > \sg_M \f{V_F-V_R}{a} $. Then, we can find  a synaptic weight distribution $H(w)$ normalized to $1$, such that \eqref{def:output} holds true for a solution of \eqref{eq:nifstst1}--\eqref{eq:nifstst4}.
\\

In the case of inhibitory signal, that means $supp  (S )\subset (-\infty, 0]$, for all $\sigma\in  \mathcal{C}^2$, the synaptic weight distribution  $H$ and $\bar N$ are unique. 
\label{thm:signalproduced}
\end{theorem}

\proof Consider an output normalized signal $S(w)\geq 0 $. Using the notations of section~\ref{sec:nonlinear}, the relations \eqref{def:output}  are reduced to building a distribution $H(w)$ normalized to $1$,  such  that  the  following relations hold
\beq
N(w) =H(w)  N_{\bar{N}, I}(w),  \qquad   H(w)= \bar{N}  \frac{S(w)}{ N_{\bar{N}, I}(w)} ,
\label{relationH}
\eeq  
and   the fixed point condition \eqref{eq:FPC} is automatically satisfied thanks to the normalization of $S$.
\\

 In other words, we look for $\bar N $ such that
$$
H(w) = \bar N  \f{Z_{\bar N,I}(w)}{a}  S(w) \quad \text{ and } \;  \int_{-\infty}^{+\infty} H(w)dw =1. 
$$
Let us mention that $H \in L^1$  because of estimate \eqref{ifststs:Zmin} and the integrability condition on $S(\cdot)$. 

These conditions are reduced to achieve the value $a$ by the mapping $ \psi : \R^+ \to \R^+$ defined by 
$$
 \psi  (\bar N) = \bar N \int_{-\infty}^{+\infty} S(w)Z_{\bar N,I}(w) dw = a.
$$
We have obviously  $\psi(0)=0$. Moreover, using  the last estimate of \eqref{ifststs:Zmin}, we obtain that 
$$
 \psi (\bar N) \geq \bar N \left(  \int_{-\infty}^{+\infty} S(w) \inf_{\bar N}  Z_{\bar N,I}(w) dw \right) \hbox{ with }  \int_{-\infty}^{+\infty} S(w) \inf_{\bar N}  Z_{\bar N,I}(w) dw >0  
$$
and so  $\psi(+\infty)=+\infty$ which implies existence of the activity $\bar{N}$ satisfying the desired nonlinearity. 
\\

In the inhibitory case, we  notice that $\psi$  is increasing as a consequence of \eqref{ifststs:Zprime} and uniqueness follows.
\\
  
The proof of Theorem~\ref{thm:signalproduced} is complete.  
\hfill $\square$

\section{Discrimination property} 
\label{sec:discrimination}

A desired property of the input-ouput map is to be able to discriminate between signals. In our language, it is to say that two different input signals $I(w)$ will generate two different network activities  $N(w)$. 
\textcolor{black}{
To state a more precise result, we need the notation, for two bounded input currents $I$ and $J$, 
$$
L(w) = e^{\f{-(\| I \|_{L^\infty} +| w|_- \sg_M )^2- (\| J \|_{L^\infty} +| w|_- \sg_M)^2}{a}} .
$$
The discrimination property  is a consequence of the functional inequality
\beq
\int | N_I(w)  - N_J(w)| dw \geq C  \int   L(w) H(w) | I(w)  - J(w)| dw ,
\label{discrim}
\eeq
for two solutions of \eqref{eq:nifstst1}--\eqref{eq:nifstst4}. The network under consideration has this property.}
\begin{theorem} [Discrimination property]  \textcolor{black}{We consider  two bounded input currents $I$ and $J$. Being given  normalized synaptic weights $H(w)$ such that $\int_{0}^{+\infty} wH(w)dw <+\infty$.  We define 
$$
\nu = \| \sg' \|_\infty \int  |w|L(w) H(w) dw <+\infty
$$
Then,  the discrimination inequality \eqref{discrim} holds true with a positive constant 
$$
C =C(\nu, a, V_F, V_R, \| I\|_\infty, \| J \|_\infty ,\sg_M).
$$ }
\label{thm:discr}
\end{theorem}
\proof \textcolor{black}{We denote by $\bar M$ a total activity obtained via Equations \eqref{eq:nifstst1}--\eqref{eq:nifstst4} stemming from the current $J$ (that is the input current is given by $J$), and  by $\bar N$ a total activity  obtained via Equations \eqref{eq:nifstst1}--\eqref{eq:nifstst4} stemming from the current $I$. }

We have 
$$
N_I(w)  - N_J(w) = H(w) [N_{\bar{N}, I}(w) -N_{\overline{M}, J}(w) ]= H(w)[ \f{a}{Z_{\bar N,I}(w)} -  \f{a}{Z_{\bar M,J}(w)} ].
$$
Therefore using the upper bound \eqref{ifststs:Zupper}, we find that there exists a constant $C$ such that 
\textcolor{black}{
\beq
|N_I(w)  - N_J(w) |= a H(w) \f { |Z_{\bar N,I}(w) - Z_{\bar M,J}(w)|} {Z_{\bar N,I}(w)  Z_{\bar M,J}(w)} \geq C L(w) H(w)  |Z_{\bar N,I}(w) - Z_{\bar M,J}(w)| .
\label{dis:est1}
\eeq }
To go further, we set $\al = I(w) + w \sg(\bar N)$, $\beta= J(w) + w \sg(\bar M)$, and write, using \eqref{eq:Z}, 
$$\bea
Z_{\bar N,I}(w) - Z_{\bar M,J}(w) &= \dis \int_{-\infty}^{V_F}  \dis \int_{v'= \max(V_R,v) }^{V_F} \left[ e^{\f{(v' - v). (v'+v -2\al)}{2a}}  - e^{\f{(v' - v). (v'+v -2\beta)}{2a}} \right]\,dv'  dv 
\\[15pt]
&= -\dis \int_{-\infty}^{V_F}  \dis \int_{v'= \max(V_R,v) }^{V_F}  \int_\al^\beta \f{v'-v}{a} e^{\f{(v' - v). (v'+v -2\gamma)}{2a}} \,d\gamma dv'  dv  .
\eea
$$
We assume, without lose of generality, that $ \beta > \alpha$.  \textcolor{black}{Then, we  have
$$\bea
\left|Z_{\bar N,I}(w) - Z_{\bar M,J}(w) \right|& \geq  \dis \int_{-\infty}^{V_F}  \dis \int_{v'=V_R }^{V_F}  \int_\al^\beta\f{v'-v}{a} e^{\f{(v' - v). (v'+v -2\gamma)}{2a}} \,dv'  dv d\gamma \
\\[15pt] & \geq \dis  \int_{-\infty}^{V_R-a}   \int_{v'=V_R }^{V_F}  \int_\al^\beta\f{v'-v}{a} e^{\f{(v' - v). (v'+v -2\gamma)}{2a}} \,dv'  dv d\gamma
\eea$$
As for all $v \in (-\infty, V_R-a)$ and $v' \in (V_R,V_F)$, it holds $\f{v'-v}{a} \geq 1$, 
 we obtain that, with 
 $$\gamma_\infty = \| I\|_\infty + \| J \|_\infty + R \sg_M ,$$
$$\bea
\left|Z_{\bar N,I}(w) - Z_{\bar M,J}(w) \right|& \geq  \dis \int_\al^\beta  \int_{-\infty}^{V_R-a}  \dis \int_{v'=V_R }^{V_F} e^\f{(v' -v)(v'+v -2 \gamma_\infty) }{2a} \,dv'  dv d\gamma 
\\[15pt] & \geq C(a, V_R, V_F , \gamma_\infty) \; (\al- \beta).
\eea$$
 We can now go back to \eqref{dis:est1} and obtain
$$\bea 
|N_I(w)  - N_J(w)|&\geq C L(w) H(w) | I(w) + w \sg(\bar N) - J(w) -   w \sg(\bar M)  | 
\\[10pt]
&\geq C L(w) H(w)\big[ | I(w)  - J(w)| - |w| \, \| \sg'\|_\infty | \bar N-  \bar M  | \big].
\eea 
$$
As a consequence, with our definition of  $\nu$, we find
$$
\int |N_I(w)  - N_J(w) | dw \geq C \int L(w) H(w) |  I(w)  - J(w)| dw - C  \nu  | \bar N-  \bar M  |
$$
and because
$$
 | \bar N-  \bar M  | = \left |\int  N_I(w) - N_J(w) dw  \right| \leq \int | N_I(w) - N_J(w) |\, dw
$$
we finally obtain
$$
\int |N_I(w)  - N_J(w) | dw \geq \f{C}{1+C  \nu}  \int L(w) H(w) |  I(w)  - J(w)| dw .
$$ 
Theorem \ref{thm:discr} is proved. }
\hfill $\square$  

\section{Synaptic weight distribution stemming from a learning rule} 
\label{sec:connectivity}

We now study how a learning rule defines a specific synaptic weight distribution.  We  assume that our network is submitted to the simplified Hebbian learning rule \eqref{hebbian} with a function $K(\cdot)$ which is  piecewise $\mathcal{C}^1$, discontinuous at $0$  and satisfies 
\beq
w K(w) >0 \hbox{ for $w \neq 0$},  \quad   0< K_m \leq  |K(\cdot)| \leq K_M \hbox{ a.e. }.
\label{sgnK}
\eeq
The sign condition is just to impose that inhibitory and excitatory neurons may change weight but remain in the same status. 
 We also give a bounded input signal. 
 Our aim here is to study possible distributions of synaptic weights generated by the pair $(K,I)$. We point out a specific difficulty which motivates the more thorough analysis in the next section,

The model we work on is the steady state equation 
\begin{equation}\label{eq:conn1}
\bepa
  \frac{\partial}{\partial v} \left[\big(- v + I(w) +w \sg(\bar N) \big) p \right]
+ \varepsilon  \frac{\partial}{\partial w} \left[\big(K(w) N(w) \bar N  - w \big) p \right]
- a \frac{\p^2 p}{\p  v^2} = N(w)  \delta(v-V_R), 
\\[5mm]
N(w) = - a \frac{\p p(V_F,w)}{\p  v},  \quad  \bar N = \dis \int_{-\infty}^{+\infty} N(w) dw, \quad H(w)= \int_{-\infty}^{V_F} p(v,w) dv, \quad \int_{-\infty}^{\infty}H(w) dw=1 , 
\eepa
\end{equation}
with the boundary conditions 
\begin{equation} \label{eq:conn2}
p(V_F,w)=0, \qquad p (-\infty,w)=0, \qquad p(v, \pm \infty) = 0.  
\end{equation}
 We  give an input signal $I(w)$ and the learning rule $K(w)$ which selects  a distribution $H$. Which are the possible synaptic weights 
$H(w)$?
\\

We show that many distributions of synaptic weights are possible and  solutions of Equation~\eqref{eq:conn1} are far from unique. We recall the definition of $Q_{\bar N,I}(w,v)$ and $N_{I,\bar N}(w)$ in Equation~\eqref{eq:linear} and state the
\begin{theorem} [Weight distribution induced by learning]
Let  $I  \in \mathcal{C}_b(\R)$ and $K$ satisfy \eqref{sgnK}. 
Then, there exists infinitely many solutions  $\overline{P}(w,v)\geq 0$ of Equation~\eqref{eq:conn1} independent of $\varepsilon$. They are given by 
$$ 
\overline{P}(v,w) = H_{ \bar{N},A}(w) Q_{\bar N,I}(w,v), \qquad \hbox{ with} \qquad  H_{ \bar{N},A}(w)=  \frac{  w  }{ {\bar N}  K(w) N_{\bar{N}, I}(w)}  \mathbb{I}_{A}(w) \geq 0, 
$$
for some appropriate  subsets $A \subset \R$  such that 
\begin{equation}\label{contrainteA}
 \int_{-\infty}^{+\infty}H_{\bar{N},A}(w)  dw =1 .
\end{equation}
\label{th:stationlearning}
\end{theorem}
\noindent \textcolor{black}{
Notice that this non-uniqueness theorem yields the question to find an organizing principle which selects the synaptic weight  among the large class built in the proof of Theorem~\ref{th:stationlearning}. This is the topic of Section \ref{sec:noise}.}
\\

\noindent{\bf Proof of Theorem  \ref{th:stationlearning}.}
The strategy of proof is as for Theorem~\ref{thm:signalproduced} and we look for a fixed point for the total activity $\bar N$ to build a solution of the nonlinear Equation~\eqref{eq:conn1}. 

   Therefore we fix a value  $\bar N>0$ and a bounded subset of $ A \subset \R$. Let the synaptic weights  $H_{\bar{N},A}(w)$ be defined by
$$
H_{ \bar{N},A}(w)=  \frac{  w  }{ {\bar N}  K(w) N_{\bar{N}, I}(w)}  \mathbb{I}_{A}(w) \geq 0 , 
$$
the sign being a consequence of the sign condition on $K$. One readily checks that, with $\overline{P}$ defined in Theorem \ref{th:stationlearning}, we have  $( N(w,t) K(w) {\bar N} - w )\overline{P} =0$ and thus $\overline{P}$ is indeed a solution of  Equation~\eqref{eq:1}.

It remains to solve the fixed point, that is to find $\bar N$ such that  the following condition holds:
$$  
 \int_{w=-\infty}^{+\infty}H_{ \bar{N},A}(w)dw =1  \quad \text{and} \quad  \bar N =  \int H_{\bar{N},A}(w) N_{I,\bar N}(w) dw.
$$
That is also written, recalling the notation \eqref{ifststs:Zdef}, 
\begin{equation}\label{fp}
  \int_{A}   \frac{w  Z_{\bar{N}, I}(w)}{ K(w)}dw= a \bar N  \quad \text{and} \quad 
  \bar N^2  =   \int_{A}   \frac{w}{ K(w)}dw  .
 \end{equation}
\textcolor{black}{Because, in Theorem  \ref{th:stationlearning}, the condition for the choice of the set $A$ is the only constraint  \eqref{contrainteA},  to conclude its proof it is enough to give a specific construction in the inhibitory case, which is addressed in  Proposition \ref{pr:stationlearning} and hence the proof of Theorem  \ref{th:stationlearning} is finished assuming Proposition   \ref{pr:stationlearning}.}
\hfill $\square$

\subsection{Solutions for learning with inhibitory  weights only.}
\label{sec:inhib}
 
 The proof of Theorem~\ref{th:stationlearning} can be concluded choosing the set $A$ so as to select inhibitory weights only. 
\begin{proposition}[Inhibitory weights]
There exists  infinitely many steady states $\overline{P}(v,w) $ of Equation~\eqref{eq:conn1} independent of $\varepsilon$ which supports in the variable $w$ are union of intervals of $(-\infty,0)$. In particular there is one of the form
$$
H_{ \bar{N}}(w)=  \frac{w }{K(w) \bar{N}  N_{\bar{N}, I}(w)} \mathbb{I}_{  -\varphi \leq w <0},
$$
\textcolor{black}{and it is unique in the case where $I'(w) \geq 0$ and where the saturation is neglected, that is $\sg(N)=N$.}
\label{pr:stationlearning}
\end{proposition}

\noindent {\bf Proof of Proposition \ref{pr:stationlearning}.} \textcolor{black}{ We first observe that if  the support of $\overline{P}(v,w) $ is equal to  $A=(-\varphi(\bar{N}), 0)$, then, with the second relation in~\eqref{fp}, necessarily $\varphi: \R^+ \to \R^+$  must be defined by}
\begin{equation}\label{def:varphi}
\bar{N}^2= \int_{- \varphi(\bar{N})}^0   - \f{w}{K(w)}dw .
\end{equation}
There exist $C_1>0$ and $C_2>0$ such that this function $\varphi$  satisfies
\beq
\textcolor{black}{\varphi(0)=0}, \qquad C_1 N  \leq \varphi(N) \leq C_2 N, \qquad C_1 \leq \varphi'(N) \leq C_2.
\label{prop_varphi}
\eeq
The first two statements are immediate and the third one follows, differentiating \eqref{def:varphi} and using \eqref{sgnK}, from the identity
\begin{equation}\label{der:varphi}
2 \bar{N}= -\varphi'(\bar{N}) \f{\varphi(\bar{N})}{K(-\varphi(\bar{N}))} \cdotp
\end{equation}
Secondly, the value $\bar N >0$ has to satisfy
\begin{equation}\label{eqfp}
\Phi(\bar N)=  \int^{0}_{-\varphi( \bar{N})}  \f{w}{K(w)} Z_{\bar{N}, I}(w) dw = a \bar{N}.
 \end{equation}
Our goal is to prove that \textcolor{black}{there exists a  positive solution of Equation \eqref{eqfp} and that it is unique when $\sg(N)=N$.} To do this, let us compute the first  two derivatives of $\Phi$.
We have 
$$\Phi'(\bar{N})= - \varphi'(\bar{N}) \f{\varphi(\bar{N})}{K(-\varphi(\bar{N}))} Z_{\bar{N}, I}(- \varphi(\bar{N}))
+  \int^{0}_{-\varphi( \bar{N})}  \f{w}{K(w)} \partial_{\bar{N}}Z_{\bar{N}, I}(w) dw.
$$
Using identity \eqref{der:varphi}, we obtain that 
$$
\Phi'(\bar{N})=  2 \bar{N} Z_{\bar{N}, I}(- \varphi(\bar{N}))+  \int^{0}_{-\varphi( \bar{N})}  \f{w}{K(w)} \partial_{\bar{N}}Z_{\bar{N}, I}(w) dw.
$$
In particular, $\Phi'(0)=0.$
Using Lemma \ref{Linearproblem}, we obtain that there exists $C>0$ such that for all $ w \leq 0$ and $\bar{N}>0$,
$$
 Z_{\bar{N}, I}(w)>C \quad \hbox{ and } \quad  \partial_{\bar{N}}Z_{\bar{N}, I}(w) \geq 0.
$$
Hence,  there exists $C>0$ such that
$$
 \Phi'(\bar{N}) \geq C \bar{N}.
$$
As $\Phi'(0)=\Phi(0)=0$, there exists at least one  \textcolor{black}{nonnegative solution to  \eqref{sgnK}}. 
\\

To prove that there is a unique one \textcolor{black}{in the case where $\sg(N)=N$}, it suffices to show that $\Phi ''>0$. 
\textcolor{black}{Using \eqref{ifststs:Zprimew}}, identity \eqref{der:varphi} and Lemma~\ref{Linearproblem}, we obtain that
\begin{align*}
 \Phi''(\bar N) = &  2 Z_{\bar{N}, I}(- \varphi (\bar{N}))  -2 \bar{N} \varphi'(\bar{N}) \partial_w Z_{\bar{N}, I}(- \varphi (\bar{N}))  + 4 \bar{N}  \partial_{\bar{N}} Z_{\bar{N}, I}(- \varphi (\bar{N}))     \\
 &+   \int^{0}_{-\varphi( \bar{N})}  \f{w}{K(w)} \partial_{\bar{N}\bar{N} }Z_{\bar{N}, I}(w) dw >0
\end{align*}
 To prove that there exists infinitely many steady states of Equation~\eqref{eq:conn1}, we notice that there exists infinitely many other  choices than $\mathbb{I}_{0\leq w \leq -\varphi( \bar N)}$   of subintervals   such that the same proof holds. An example is, given $w_0 \leq 0$,  to consider the function $\widetilde{\varphi}(\bar N)$ such that 
$$
H_{ \bar{N}}(w)=  \frac{-w }{ \bar{N}  N_{\bar{N}, I}(w)} \mathbb{I}_{   w_0 \geq w \geq -\widetilde{\varphi}(\bar{N})}, \qquad N(w)= N_{\bar{N}, I}(w) H_{ \bar{N}}(w)
$$
where  $ \widetilde{\varphi}(\bar{N})$  is the value determined by 
$$
\int_{-\infty}^0 H_{ \bar{N}}(w) dw=1. 
$$
The above argument applies directly and this concludes the proof of  Proposition~\ref{pr:stationlearning} and of Theorem~\ref{th:stationlearning}. 
\hfill $\square$

\subsection{Non existence result for  learning with excitatory  weights only} 
\label{sec:exc}

One might try the same approach and try to find purely excitatory weights. This is not always possible and we have the
\begin{proposition}[Excitatory weights]
We take  for $w>0$  a bounded signal $I$. When $\sg(N)=\sg_0 \f{N}{1+N}$ with $\sg_0$  large enough, there is no solution of~\eqref{eq:conn1} with a weight distribution  under the form 
\textcolor{black}{
$$
H_{ \bar{N},A}(w)=  \frac{  w  }{ {\bar N}  K(w) N_{\bar{N}, I}(w)}  \mathbb{I}_{A}(w),\qquad \text{with}ส\quad A=(0,\varphi ).
$$}
\label{pr:excitatory}
\end{proposition}
\textcolor{black}{ This Proposition implies that, for the purely excitatory case, we may not have convergence of the solution of Equation ~\eqref{eq:conn1} to a stationary state. As an example, in the situation of Proposition \ref{pr:excitatory}, to hope having convergence to a stationary state, we have to deal with an initial condition where the support of $H$  and $(-\infty,0)$ is non empty. }

\medskip

\proof  Using the same proof as for Proposition~\ref{pr:stationlearning}, we impose, still because of the conditions in \eqref{fp}, 
$$
{\bar N}^2= \int_{0}^{ \varphi(\bar N)}   \frac{w}{K(w)} dw.
$$
and the properties~\eqref{prop_varphi} still hold.  According to the other condition in \eqref{fp}, we examine the condition
$$
a\bar{N}=  \int_{0}^{\varphi(\bar N)}   \frac{w}{K(w)} Z_{\bar{N}, I}( w) dw : =\Phi(\bar N).
$$
We notice that 
$$
\Phi'(\bar{N}) =  \varphi'(\bar{N}) \f{\varphi(\bar{N})}{K(\varphi(\bar{N}))} Z_{\bar{N}, I}( \varphi(\bar{N})) + \int_{0}^{\bar N^2}  \p_{\bar{N}} Z_{\bar{N}, I}( w) dw .
$$
Since $\Phi'(0)=0$, we expect that, if there was a fixed point $N_0$, then the first fixed point will be such that $\Phi'(N_0) \geq 1$. However at such a fixed point, we find, because $\p_{\bar{N}} Z_{\bar{N}, I}(w)  <0$ for $w>0$, 
$$
\Phi'(N_0) \leq   \varphi'(Nยก0) \f{\varphi(N_0)}{K_M}   \dis \int_{-\infty}^{V_F}  \dis \int_{v'= \max(V_R,v) }^{V_F} e^{\f{(v' - v). [v'+v +2\|I \|_\infty-2 \sg_0 \f{N_0 \varphi{(N_0^2)}}{1+ N_0})]}{2a}} \,dv'  dv.
$$
From the properties~\eqref{prop_varphi}, we conclude that  for $ \sg_0$ large enough, we have necessary
$$
\Phi'(N_0)< 1 ,
$$ 
which is a contradiction and concludes the proof. 
\hfill $\square$

\section{Selection of inhibitory synaptic weights by noise} 
\label{sec:noise}

In this section, \textcolor{black}{we choose $K(w)=-1$ for $w \leq 0$ and we only consider inhibitory interconnections. In view of the result of Section~\ref{sec:connectivity}, we try to find a selection principle for the synaptic weight distribution which would single out the choice $A=(-\vp,0)$ established in Proposition~\ref{pr:stationlearning}. Indeed, among the infinitely many steady states constructed in Theorem~\ref{th:stationlearning}, numerical evidence, in Section~\ref{sec:LTPR}, indicates that  a unique stationary state is selected  with   $A= (-\sqrt{ 2}\;  \bar{N},0)$.
}

Two difficulties occur, namely the selection of the set $A$ and the uniqueness of the value $\bar N$ when solving the fixed point~\eqref{fp}. 
 As in the inhibitory case, blow-up does not occur for Leaky Integrate and Fire models \cite{CPSS}, we simply assume that $\sigma \big(\bar N(t) \big)=\bar N(t)$.

A possible organizational principle can be noise, which is compatible with numerical diffusion in the observations of Section~\ref{sec:connectivity}.  Therefore, we heuristically study the stationary state of a modified equation with a Gaussian noise, of  intensity $\nu >0$, on the variable $w$.  We use slow-fast limit in order to take into account that learning is on a slow scale compared to neural activity. Then, we may compute more easily the potential stationary states of  the new equation given by 
  \begin{equation}\label{eq:Slow-Fast}
  \bea
\e \frac{\partial p_{\epsilon}}{\partial t} + \frac{\partial}{\partial v} \left[\big(- v + I(w) + w \bar N_{\epsilon}(t) \big) p_{\epsilon} \right]
+ & \e \frac{\partial}{\partial w} \left[\big( - N_{\epsilon}(w,t) \bar N_{\epsilon}(t)  - w \big) p_{\epsilon} \right]
\\[10pt]
&- a \frac{\p^2 p_{\epsilon}}{\p  v^2} - \varepsilon \nu  \frac{\p^2 p_{\epsilon}}{\p  w^2} = N_{\epsilon}(w,t)  \delta(v-V_R), 
\eea
 \end{equation}
 with  boundary conditions  adapted to the purpose of dealing only with $w \leq 0$,
\begin{equation} \label{eq:noise2}
p_{\epsilon}(V_F,w,t)=0, \qquad p_{\epsilon} (-\infty,w,t)=0, \qquad \left(  N_{\epsilon}(w,t) \bar{N}_{\epsilon}(t) +w\right)  p_{\epsilon} = -	 \nu  \frac{\partial p_{\epsilon}}{\partial w}, \quad \text{at }\; w=0.
\end{equation}
Here $\varepsilon >0$ represents the time scale of learning and thus vanishes in the limit of fast network adaptation vs slow learning. 
\\

 \textcolor{black}{In a first and formal step in our analysis, we consider the fast time scale $\e \to 0$. This yields  the steady state Integrate and Fire  density distribution as studied in Section~\ref{sec:nonlinear}. Since the synaptic weight distribution takes a value $\widetilde H(w)$ that changes according to the slow time scale, we  fix it here  and find }
$$
p^*[\widetilde H] = \widetilde H(w) Q_{\bar N[\widetilde H],I}(v,w)
$$
with $Q_{\bar N,I}$ defined through \eqref{eq:linear}, \eqref{eq:FPC}.
Then,  $\bar N[\widetilde H]$ is solution of the fixed point equation 
$$
 \bar N[\widetilde H] = \int_0^\infty \widetilde H(w) N_{\bar N[\widetilde H],I}(w) dw
 $$
 with $N_{\bar N[\widetilde H],I}$ is defined by Equation~\eqref{ifststs:Zdef}.  

 \textcolor{black}{In a second step, we can integrate in $v$  Equation~\eqref{eq:Slow-Fast} and divide by $\e$. Recalling that, from~\eqref{eq:connectivity_distr}, $\widetilde H_{\epsilon}(w,t)= \int_{-\infty}^{V_F} p_{\epsilon}(v,w,t)dv$, we obtain
$$
\frac{\partial \widetilde{H}_{\epsilon}}{\partial t} + \frac{\partial }{\partial w}\left(\left(  -\widetilde{N}_{\epsilon} (w) \bar{N}_{\epsilon}(t) - w\right)      \widetilde{H}_{\epsilon}	    \right)= \nu \frac{\partial^2 \widetilde{H}_{\epsilon}}{\partial w^2} .
$$
With the equilibrium of the first step, we find the limit 
\begin{equation}
\frac{\partial \widetilde{H}}{\partial t} + \frac{\partial }{\partial w}\left(\left(  -\widetilde{N} (w) \bar{N}(t) - w\right)      \widetilde{H}    \right)= \nu \frac{\partial^2 \widetilde{H}}{\partial w^2} , \quad w \leq 0
\label{eq:hbarnoise}
\end{equation}
}
where  
\begin{equation}
\widetilde{N}(w,t) = N_{\bar{N}[\widetilde H(.,t)],I} \widetilde H(w,t), \quad  N_{\bar{N}[\widetilde H(.,t)],I}(w,t):= -a\frac{\p}{\p v}Q_{\bar{N}[\widetilde  		H(.,t)]}(V_F,w), \quad \bar{N}(t)= \bar{N}[\widetilde H(.,t)],
\label{lm:connectivity}	
\end{equation}
and with the no-flux boundary condition 
\begin{equation}
\left(  \widetilde{N}(w,t) \bar{N}(t) +w\right)   \widetilde{H} = -	 \nu  \frac{\partial \widetilde{H}}{\partial w}, \quad \text{at }\; w=0.
\label{lm:noiseBC}	
\end{equation}

Notice that the form  of $ \widetilde{N}(w,t)$ makes that Equation~\eqref{eq:hbarnoise} is nonlinear hyperbolic, closely related to first order scalar conservation laws, \cite{Dafermos, SerreBook}. Therefore, we may expect that discontinuities (shocks) can be formed and that noise selects indeed a specific solution, namely the entropy solution. This is stated in the following Theorem: 
%
\begin{theorem} [Small noise limit] Assume that   $I'(w) \geq 0$. As $\nu \to 0$, the steady state of  \textcolor{black}{Equation~\eqref{eq:hbarnoise}-- \eqref{lm:noiseBC} converges to  the   unique steady state built in Proposition~\ref{pr:stationlearning}} and supported by the single interval $[-A,0]$. \label{th:noise}
\end{theorem}
%
 
\proof
After integrating the equation for the  steady states of~\eqref{eq:hbarnoise}, and using the boundary condition at $w=0$, we find that each stationary state $\widetilde H_\nu$ satisfies
\begin{equation}
\left(  \widetilde{H}_\nu N_{\bar{N}[\widetilde H_\nu],I}(w) \bar{N} + w\right)      \widetilde{H}_\nu	= - \nu  \frac{d \widetilde{H}_\nu}{dw}.
\label{lm:noiseWTH}	
\end{equation}
\textcolor{black}{ We first observe that solutions $\widetilde{H}_\nu$ of such an equation cannot vanish at a point  because  they are given by an exponential.
\newline
Next, we claim that there is $w_\nu <0$ such that
\begin{equation}
 \widetilde{H}_\nu' (w) >0 \text{ for } w<w_\nu, \ \hbox{ and }  \ \widetilde{H}_\nu' (w) < 0 \text{ for } w > w_\nu .
\label{lm:noiseMD}	
\end{equation} 
Indeed, since  $ \widetilde{H}_\nu (0)>0$,  for $w$ close to zero, we have $\widetilde{H}_\nu N_{\bar{N}[\widetilde H_\nu],I}(w) \bar{N} + w >0$ and thus  $\widetilde{H}_\nu' (w) < 0$. Because $ \widetilde{H}_\nu$ is integrable on $(-\infty,0)$, there has to be a largest value $w_\nu$ where  
$$0= \widetilde{H}_\nu N_{\bar{N}[\widetilde H_\nu],I}(w_\nu) \bar{N} + w_\nu \  \hbox{ and }  \ 0=  \widetilde{H}_\nu' (w_\nu).$$
Finally, on $(-\infty, w_\nu)$ we necessarily have $\widetilde{H}_\nu' (w) >0$ because $\f{-w}{N_{\bar{N}[\widetilde H_\nu],I}(w)}$ is a decreasing function thanks to the assumption $I'(w) >0$ which implies $N_{\bar{N}[\widetilde H_\nu],I}'>0$ using \eqref{ifststs:Zprimew} because $N_{\bar{N}[\widetilde H_\nu],I} = \frac{a}{Z_{\bar{N}[\widetilde H_\nu],I}}$.
Therefore, if there was a second crossing point $w_1<w_\nu$, where $\widetilde{H}_\nu N_{\bar{N}[\widetilde H_\nu],I}(w_1) \bar{N} =w_1$, we should have  both $\widetilde{H}_\nu' (w_1) < 0$ (to cross a decreasing function)  and, the condition $\widetilde{H}_\nu' (w_1) = 0$ from \eqref{lm:noiseWTH}. A contradiction which states \eqref{lm:noiseMD}.
\newline
From this property, that  $\int_{-\infty}^0 \widetilde H_\nu(w) dw=1$ and the control from below and above of $N_{\bar{N}[\widetilde H_\nu],I}(w)$ using Lemma~\ref{Linearproblem}, we conclude that $ \widetilde{H}_\nu $ is uniformly bounded and has the uniform decay $\exp(-\f{w^2}{2 \nu })$ as $w\to -\infty$. Therefore we may pass to the limit in~\eqref{lm:noiseWTH} and conclude that the limit $\wt H$ satisfies either $\wt H (w)=0$, or $\wt H (w) N_{\bar{N}[\widetilde H],I}(w) \bar{N} + w =0$. From~\eqref{lm:noiseMD}, this identifies the support of $\wt H$ as stated in Theorem~\ref{th:noise}.}
\hfill $\square$  

\section{Learning, testing and pattern recognition}
\label{sec:LTPR}

 \textcolor{black}{Based on numerical simulations, we illustrate the discrimination property stated in Section~\ref{sec:discrimination}.} We consider the following two-phase setting:
\\[1mm] 
\noindent \textbf{Learning phase}
\begin{enumerate}
\item An heterogeneous input $I(w)$ is presented to the system, while the learning process is active. \textcolor{black}{
 The chosen initial data is supported on inhibitory weights so as to avoid the complexity of excitatory cases and the learning rule is determined for the inhibitory weights  by $ -N(w) \bar N$, as in section~\ref{sec:connectivity}, by taking $K(w)=-1$ if $w \leq 0$.}
\item After some time, the synaptic weight distribution $H(w,t)$ converges to an equilibrium distribution $H^*_I(w)$, which depends on $I$.
\end{enumerate}
\textbf{Testing phase}
\begin{enumerate}
\item The learning process is now switched off, and a new input $J(w)$ is presented to the system.
\item After some time, the solution $p_J(v,w,t)$ reaches an equilibrium $p^*_J(v,w)$, which can be summarized by the output signal $N^*_J(w)$ which is the neural activity distribution across the heterogeneous populations.
\end{enumerate}

\textcolor{black}{The numerics has been performed using a finite difference method. For the Fokker-Planck equation on the potential, we use the Sharfetter-Gummel method \cite{Jungel}. For the transport equation on the weight variable, we use an upwind scheme \cite{Bouchut, Leveque}. The matlab code is available on demand to one of the authors.}
\\

Then, from the mathematical analysis performed in previous sections, we know that the following "pattern recognition" property will be observed: \textit{the system can detect whether the new input $J(w)$ is actually the same one that has been presented during the learning phase, i.e. $I(w)$: indeed, in this case, $N^*_J(w) = w \mathbf{1}_{[-A,0]}$ has a very specific shape.} A remarkable feature is that this specific shape does not depend upon the original input $I$ that has been learned in the learning phase: it is an intrinsic property of the system. This is particularly interesting because it implies that detecting a learned pattern could be implemented by an external system which would be independent of the given pattern.
\\

\begin{figure}[t]
\begin{center}
\includegraphics[width=7cm]{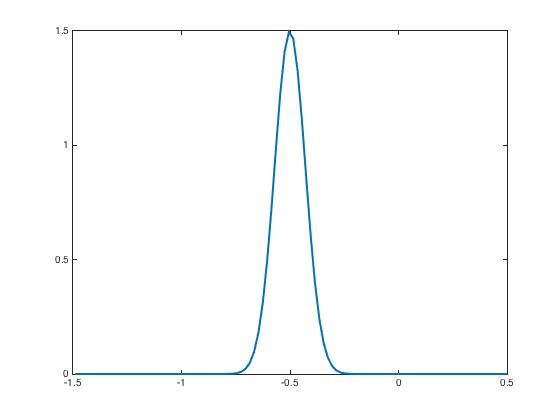} 
\includegraphics[width=7cm]{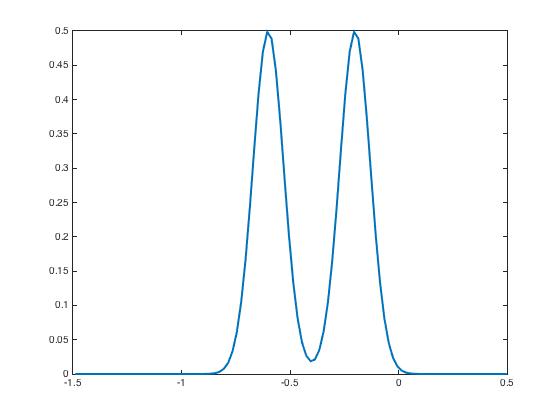} 
\end{center}
\vspace{-11mm}
\caption{(Two input signals) We have used two input signals that we denote $I(w)$ on the left and $J(w)$ on the right.}
\label{fig:input}
\end{figure}

\begin{figure}[t]
\begin{center}
\includegraphics[width=7cm]{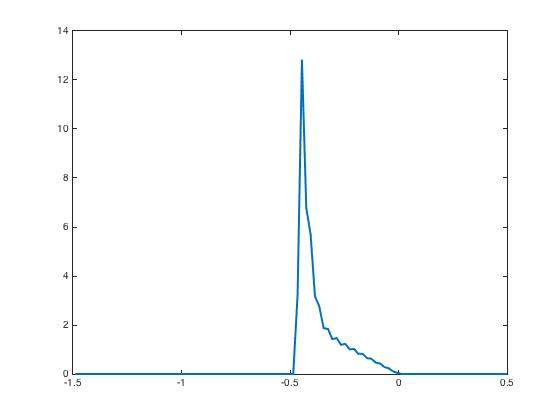} 
\includegraphics[width=7cm]{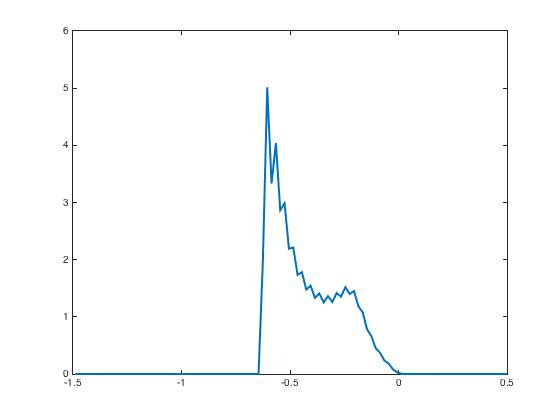} 
\end{center}
\vspace{-11mm}
\caption{(Two connectivities) This figure displays the synaptic connectivities obtained after learning with the input signals of Figure~\ref{fig:input}.}
\label{fig:connect}
\end{figure}

\begin{figure}[t]
\begin{center}
\includegraphics[width=7cm]{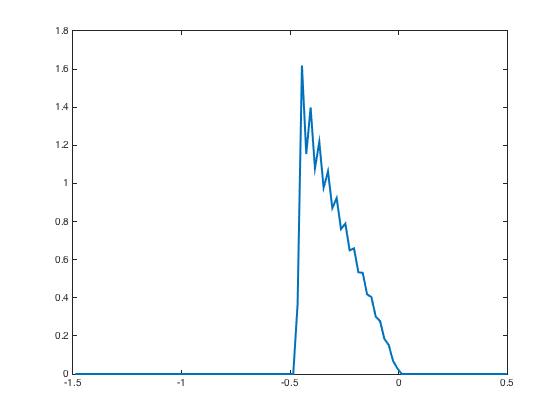} 
\includegraphics[width=7cm]{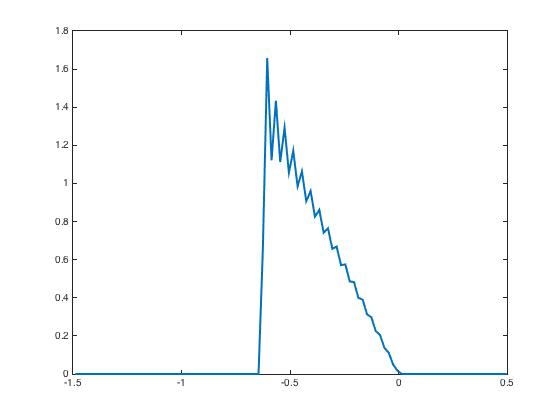} 
\end{center}
\vspace{-11mm}
\caption{(Output with learned signal) This figure displays the netwotk activities $N(w)$  when learned with the two input signals of Figure~\ref{fig:input},  $I$ on the left, $J$ on the right.}
\label{fig:output1}
\end{figure}

\begin{figure}[t]
\begin{center}
\includegraphics[width=7cm]{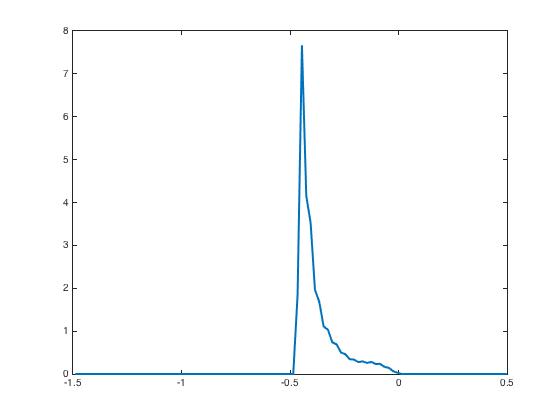} 
\includegraphics[width=7cm, height=5.5cm]{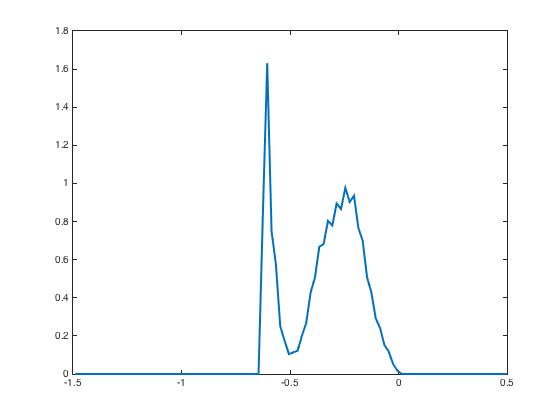} 
\end{center}
\vspace{-11mm}
\caption{(Output with the other signal) With the synaptic weights of the learned signal $I$, we display the activty for the input signal $J$ (left), and for learned signal $J$ we display  the activty for the input signal $I$ (left).}
\label{fig:output2}
\end{figure}

\textcolor{black}{To illustrate this pattern recognition property we display in Figure~\ref{fig:input} the two input signals we have used for  the learning-testing set-up
$$
I(w)= 1.5*\exp\left(-\f{(w+0.5)^2}{0.01}\right), \quad J(w)= .5*\exp\left(-\f{(w+0.2)^2}{0.01}\right)+.5*\exp\left(-\f{(w+0.6)^2}{0.01}\right) . 
$$
After presentation of these input currents $I(w)$ and $J(w)$, the synaptic weight distribution converges to $H^*_I$ that are displayed in Figure~\ref{fig:connect}. The corresponding network activity $N(w)$ are shown in Figure~\ref {fig:output1}.
\\
During the testing phase, learning is off and the system reacts differently according to the input it receives: if the new input is the same as the learned one, then the neural activity distributes according to the specific shape predicted by the theory and already shown in Figure~\ref{fig:output1}, indicating that the network has recognized the  learned pattern.  Whereas if the new input is not the same, here we invert $I$ and $J$ as input currents, then the neural activity distributions have a very different shape Figure~\ref{fig:output2}. This illustrates the discrimination property.
}

\section{Conclusions and perspectives} 
\label{sec:conclusion}

We have introduced a novel mathematical framework to study learning mechanisms in macroscopic models of spiking neuronal networks by considering plasticity between neural subpopulation and the overall mean-field activity. When ignoring the learning rule, we have characterized the synaptic weight distribution which generates a given output signal, and we have shown a discrimination property. When the learning rule is activated, we have studied the multiple synaptic weight equilibria of the global coupled system with learning. A selection by noise selects a unique equilibria which is also observed numerically.  Furthermore, we have investigated the ability of such models to perform pattern recognition tasks.
\\

The class of models studied in this article are subject to several limitations and mainly that the network is coupled via a global activity and not by pairwise interactions. A related limitation is that stability and convergence to a unique equilibrium point depend  on the excitatory/inhibitory nature of the synaptic weight as it does for the noisy integrate-and-fire network model.  Because we have targeted mathematically proved results, we had to assume that the input signal is time independent, which is a restriction in the theory.
\\

 To further extend our study, one should investigate other learning rules. A possible extension is to use pairwise connections, leading to the following extension of our system
 \begin{equation}\label{eq:persp} \left\{\begin{array}{l}
\frac{\partial p}{\partial t} + \frac{\partial}{\partial v} \left[\big(- v + I(w) + \int_{-\infty}^\infty  C(w,w',t) N(w',t)dw'\big) p \right]
- a \frac{\p^2 p}{\p  v^2} = N(w,t)  \delta(v-V_R), 
\\[5pt]
N(w,t) : =  -a \frac{\p p}{\p v} (V_F,w,t) \geq 0, \qquad \bar N(t)= \dis \int_{-\infty}^\infty  N(w,t) dw,
\end{array} \right.
\end{equation}
$$
\frac{\partial}{\partial t} C(w,w',t) =  K(w,w') N(w,t) N(w',t) - C(w,w',t) .
$$
  In closer connection with biological mechanisms such as spike-timing dependent plasticity, which may also be integrated in the model with convolution operators. Other models of neuronal dynamics, beyond spiking models, such as rate models or coupled oscillator systems, could also be studied and compared within the proposed formalism.
\\

 Finally, to make the link with the fields of pattern recognition and machine learning deeper, further questions can be considered, for instance to quantify the discrimination ability between two signals or to evaluate the number and complexity of attractors, possibly dynamic, which can be stored into the synaptic weight distribution. 
 \\
 \\
 
\noindent {\bf Acknowledgment:}  BP and DS are supported by the french "ANR blanche" project Kibord:  ANR-13-BS01-0004.
%
%
%

\bibliographystyle{siam}
\bibliography{BibPSW}

\end{document}